\documentclass[aps,prl,onecolumn,showpacs,preprintnumbers]{revtex4}
\usepackage{amsmath}
\usepackage{dcolumn}
\usepackage{bm}
\usepackage{subfigure}
\usepackage{amsfonts}
\usepackage{amssymb}
\usepackage{makeidx}
\usepackage{epsfig}
\usepackage{graphicx}

\begin{document}

\title{Magnification effect of Kerr metric by configurations of
collisionless particles in non-isotropic kinetic equilibria}
\author{Claudio Cremaschini}
\affiliation{Institute of Physics and Research Center for Theoretical Physics and
Astrophysics, Faculty of Philosophy and Science, Silesian University in
Opava, Bezru\v{c}ovo n\'{a}m.13, CZ-74601 Opava, Czech Republic}
\author{Zden\v{e}k Stuchl\'{\i}k}
\affiliation{Institute of Physics and Research Center for Theoretical Physics and
Astrophysics, Faculty of Philosophy and Science, Silesian University in
Opava, Bezru\v{c}ovo n\'{a}m.13, CZ-74601 Opava, Czech Republic}
\date{\today }

\begin{abstract}
A test fluid composed of relativistic collisionless neutral particles in the
background of Kerr metric is expected to generate non-isotropic equilibrium
configurations in which the corresponding stress-energy tensor exhibits
pressure and temperature anisotropies. This arises as a consequence of the
constraints placed on single-particle dynamics by Killing tensor symmetries,
leading to a peculiar non-Maxwellian functional form of the kinetic
distribution function describing the continuum system.

Based on this outcome, in this paper the generation of Kerr-like metric by
collisionless $N-$body systems of neutral matter orbiting in the field of a
rotating black hole is reported. The result is obtained in the framework of
covariant kinetic theory by solving the Einstein equations in terms of an
analytical perturbative treatment whereby the gravitational field is
decomposed as a prescribed background metric tensor described by the Kerr
solution plus a self-field correction. The latter one is generated by the
uncharged fluid at equilibrium and satisfies the linearized Einstein
equations having the non-isotropic stress-energy tensor as source term. It
is shown that the resulting self-metric is again of Kerr type, providing a
mechanism of magnification of the background metric tensor and its
qualitative features.
\end{abstract}

\maketitle


\textbf{Keywords:} Kerr metric; Killing tensor symmetry; Kinetic
equilibria; Non-isotropic equilibria

\section{1 - Introduction}

Collisionless $N-$body systems formed by a large number of constituent
particles, namely in which $N\gg 1$, and composed by neutral and/or charged
matter are expected to arise in several astrophysical environments
associated with strong gravitational and electromagnetic (EM) fields.
Examples include the case of collisionless gas clouds \cite%
{gas0,gas00,gas1,gas2,gas3,gas4}, dark matter (DM)\ halos \cite%
{DM1,DM3,DM4,DM5,DM6,DM7,DM8} around stellar-mass or galactic-center black
holes, collisionless stellar systems belonging to globular clusters or
orbiting central galactic regions \cite{gravi,r85,star0,star1} as well as
magnetized plasmas in the surrounding of compact objects giving rise to
accretion-disc systems, relativistic jets and hot coronal structures \cite%
{coll-2011,APJS,coll-2014}. In these scenarios space-time curvature effects
can be relevant, while both single-particle and macroscopic matter fluid
velocities can become relativistic, leading to the occurrence of
corresponding relativistic regimes. Under these circumstances, the
statistical description of the dynamical and thermodynamical features of
collisionless systems must be reached in terms of covariant microscopic and
Vlasov kinetic theories, to be generally supplemented by both Maxwell and
Einstein equations \cite{gravi,degroot,PoP2014-1,PoP2014-2}. Within the
kinetic description, the fundamental quantity is represented by the kinetic
distribution function (KDF) $f$ in terms of which statistical averages and
corresponding observable fluid fields can be evaluated. The KDF is defined
in the single-particle phase-space and its dynamical evolution in the
collisionless regime is determined by the Vlasov equation. In Lagrangian
representation, the latter takes the general form%
\begin{equation}
\frac{d}{ds}f\left( \mathbf{x}\left( s\right) ,s\right) =0,  \label{vlasov-1}
\end{equation}%
where\ $s$ and $\mathbf{x}\left( s\right) \equiv \left( r^{\mu }\left(
s\right) ,u^{\mu }\left( s\right) \equiv \frac{dr^{\mu }\left( s\right) }{ds}%
\right) $ denote respectively the single-particle\ world-line proper time
and phase state, being $r^{\mu }$ and $u^{\mu }$ the particle $4-$position
and velocity $4-$vector. The kinetic description provides\ at the same time
also a consistent framework for determining related continuum fluid
theories, i.e., relativistic hydrodynamic \cite{ros,mano2,rezhd2} and/or
magnetohydrodynamic \cite{rezmhd,ma-2,ma-3,ma-4,ma-5,ma-6} treatments. In
fact, velocity-integrals of the KDF define appropriate fluid fields, namely
physical observables (e.g., number density and temperature), while velocity
moments of the kinetic equation (\ref{vlasov-1}) determine the corresponding
set of continuum fluid equations. In this sense, the kinetic treatment
provides a fundamental level of statistical description of $N-$body systems
and is the approach followed in this paper.

A remarkable property of collisionless $N-$body systems is that they can
generate phase-space anisotropies and are therefore generally characterized
by the occurrence of non-isotropic KDFs. In order to define properly such a
concept, we consider the case of a single-species system composed of neutral
particles having unitary rest mass, so that $m=1$. Along the particle
geodesic trajectory the line element $ds$ satisfies the identity
\begin{equation}
ds^{2}=g_{\mu \nu }\left( r^{\alpha }\right) dr^{\mu }dr^{\nu },
\end{equation}%
where $g_{\mu \nu }\left( r^{\alpha }\right) \in \mathcal{R}^{4}$ is a
generic position-dependent metric tensor with signature $\left(
-,+,+,+\right) $, while $\mathcal{R}^{4}$ is a $4-$dimensional smooth
Lorentzian manifold. Correspondingly, the $4-$velocity $u^{\mu }$ satisfies
the mass-shell normalization $g_{\mu \nu }\left( r^{\alpha }\right) u^{\mu
}u^{\nu }=1$. For single-particle dynamics the unconstrained 8-dimensional
phase-space $\Omega $ is then defined as $\Omega =\mathcal{R}^{4}\times
\mathcal{V}^{4}$, where the 4-dimensional velocity space $\mathcal{V}^{4}$
is defined as the tangent bundle of $\mathcal{R}^{4}$. The corresponding
constrained velocity space whose 4-vectors are subject to the mass-shell
normalization is indicated symbolically with $\mathcal{V}^{3}$.

Following the approach developed in Refs.\cite{PoP2014-2,Bek1,Bek2}, we
introduce a set of orthogonal unit $4-$vectors $\left( a^{\mu },b^{\mu
},c^{\mu },d^{\mu }\right) $ to be related to the coordinate system, where $%
a^{\mu }$ and $\left( b^{\mu },c^{\mu },d^{\mu }\right) $ are respectively
time-like and space-like. This implies the decomposition of the particle $4-$%
velocity as%
\begin{equation}
u^{\mu }\equiv u_{0}a^{\mu }+u_{1}b^{\mu }+u_{2}c^{\mu }+u_{3}d^{\mu },
\label{4-veeldeco}
\end{equation}%
to be denoted as tetrad representation, where $\left(
u_{0},u_{1},u_{2},u_{3}\right) $ are the corresponding components which are
separately $4-$scalars. This\ representation yields%
\begin{equation}
u_{0}=\sqrt{u_{1}^{2}+u_{2}^{2}+u_{3}^{2}-1},  \label{u-zero}
\end{equation}%
which expresses the time-component of the 4-velocity with respect to the
space-components. Based on these premises, in the following a relativistic
KDF $f$ is said to be isotropic on the velocity space $\mathcal{V}^{3}$ if
it carries even powers of all the velocity components $\left(
u_{1},u_{2},u_{3}\right) $ and this functional dependence is isotropic. A
particular case of isotropic dependence is through the velocity component $%
u_{0}$ only, as it is exemplified by the relativistic Maxwellian function $%
f_{M}$ \cite{degroot}. In contrast, a KDF $f$ is said to be non-isotropic on
$\mathcal{V}^{3}$ if it exhibits a non-isotropic dependence on the even
powers of the same components $\left( u_{1},u_{2},u_{3}\right) $. Thanks to
the tetrad representation, these definitions apply necessarily also to KDFs
associated with non-relativistic collisionless regimes.

A basic problem to investigate which is related to the occurrence of
phase-space anisotropies and the statistical behavior of collisionless
systems is the existence of equilibrium states. In the Vlasov theory,
kinetic equilibria are realized by smooth and strictly-positive KDFs which
are summable on the velocity space $\mathcal{V}^{4}$ and depend only on a
suitable set of invariants $I(\mathbf{x})$ which satisfy the condition $%
\frac{d}{ds}I(\mathbf{x}(s))=0$. More precisely, the real scalar functions $%
I(\mathbf{x})$\ identify the so-called integrals of motion, i.e., invariant
phase functions which do not explicitly depend on $s$ \cite{wald}. The set $%
I(\mathbf{x})$ can include for example the particle energy and/or the
particle canonical momentum when respectively the conditions of time and/or
space symmetries apply to particle dynamics, with respect to a given
coordinate system. Given the set $I(\mathbf{x})$ characteristic of the
problem to be studied, kinetic equilibria can be realized by requiring a
functional dependence for the equilibrium KDF of the type $f=f\left( I(%
\mathbf{x}(s))\right) $. This technique is referred to as the method of
invariants and has been investigated in depth in previous works, in
particular in Refs.\cite{Cr2010,Cr2011,Cr2011a} for non-relativistic
systems, in Refs.\cite{PoP2014-1,PoP2014-2}\textbf{\ }for plasmas in curved
space-time and in Ref.\cite{ijmpd-2016} for test fluid of neutral matter
orbiting around a rotating black hole. It has been shown that anisotropy
effects can arise either:\newline
A) Due to relativistic single-particle dynamics and related conservation
laws \cite{p-1,p-2,p-3,p-4,p-5}, including the occurrence of adiabatic
invariants \cite{p-0}. For charged particles in electromagnetic fields these
features are enriched by gyrokinetic dynamics in plasmas \cite%
{PoP2014-1,PoP2014-2,Bek1,Bek2} and plasma kinetic regimes \cite{Cr2012},
taking into account the constraints placed by non-uniform magnetic fields on
particle dynamics and confinement mechanisms.\newline
B) Due to the peculiar form of the KDF itself, which may depart from an
isotropic local Maxwellian function \cite{Cr2010,Cr2011}, implying the
occurrence of collective phenomena which are distinctive of the statistical
behavior of collisionless $N-$body systems.

The occurrence of phase-space anisotropies represents a relevant feature of
equilibrium configurations of collisionless systems, since non-isotropic
solutions exhibit peculiar physical properties. Here we refer in particular
to the observable velocity and temperature/pressure profiles of
astrophysical systems. Concerning the first issue, phase-space anisotropies
of kinetic origin can give rise to collective drift-velocities, namely fluid
velocities along characteristic spatial directions which affect the bulk
motion of the continuum fluid introducing drift corrections. In the case of
plasmas different mechanisms have been identified which are responsible for
the occurrence of such macroscopic drift motions. The most relevant ones
are:\ 1) those associated with conservation of particle canonical momentum
in axisymmetric systems in the presence of strong magnetic fields, i.e. the
so-called diamagnetic effects in strongly-magnetized plasmas; 2)\ those
associated with Larmor rotation of charges and the related confinement
mechanism for particle trajectories approximating magnetic field lines in
magnetized plasmas, usually denoted as finite Larmor-radius (FLR) effects
\cite{Cr2013}. This represents the most efficient mechanism for the
generation of phase-space anisotropy in magnetized plasmas and yields the
bi-Maxwellian character of the equilibrium KDF;\ 3)\ those associated with
energy conservation in stationary or slowly time-varying kinetic equilibria
\cite{APJS,Cr2010,Cr2011}. In addition, phase-space anisotropies can
generate also shear-flow phenomena characterized by strong gradients of
angular velocity in axisymmetric systems \cite{Cr2013c} as well as the
phenomenon of kinetic dynamo, namely magnetic field generation by local
current densities in plasmas \cite{coll-2011,Cr2011a}. Finally, additional
contributions to kinetic anisotropies in plasmas can originate also in the
framework of energy-independent kinetic\ equilibria \cite{PoP2014-1,PRE-new}
and in axisymmetric systems for particles in epicyclic motion around minima
of effective potential \cite{Cr2013b}.

On similar grounds, phase-space anisotropies can affect the physical
temperature and pressure profiles of collisionless systems, leading to the
occurrence of the phenomenon of temperature and pressure anisotropies. For
relativistic systems this can be best understood by investigating the
properties of the corresponding system stress-energy tensor $T^{\mu \nu
}\left( r^{\alpha }\right) $ \cite{degroot}. This is defined by the $4-$%
velocity integral over the KDF $f$ as%
\begin{equation}
T^{\mu \nu }\left( r^{\alpha }\right) =2\int_{\mathcal{V}^{4}}\sqrt{-g}%
d^{4}u\Theta \left( u_{0}\right) \delta \left( u^{\mu }u_{\mu }-1\right)
u^{\mu }u^{\nu }f,  \label{tmunu}
\end{equation}%
where $\sqrt{-g}$ is the square-root of the determinant of the metric
tensor. Invoking the tetrad representation for the $4-$velocity and upon
performing the integration using the Dirac-delta, one obtains the equivalent
representation
\begin{equation}
T^{\mu \nu }\left( r^{\alpha }\right) =\int_{\mathcal{V}^{3}}\frac{\sqrt{-g}%
d^{3}u}{\sqrt{u_{1}^{2}+u_{2}^{2}+u_{3}^{2}-1}}u^{\mu }u^{\nu }f,
\label{tmunu-bis}
\end{equation}%
where the component $u_{0}$ of the $4-$velocity is intended as dependent on
the other components according to Eq.(\ref{u-zero}). According to the
definitions introduced above, a stress-energy tensor $T^{\mu \nu }\left(
r^{\alpha }\right) $ is said to be isotropic if it is defined in terms of an
isotropic KDF $f$. An example is the stress-energy tensor associated with a
relativistic Maxwellian KDF $f_{M}$ which describes ideal fluids, to be
denoted as $T_{M}^{\mu \nu }\left( r^{\alpha }\right) $ \cite{degroot}.
Correspondingly, the same tensor field $T^{\mu \nu }\left( r^{\alpha
}\right) $ will be referred to as non-isotropic stress-energy tensor if the
KDF in Eq.(\ref{tmunu-bis}) is a non-isotropic distribution. The properties
of non-isotropic stress-energy tensors for both plasmas and uncharged matter
will be detailed below in separate sections.

On general grounds, in the case of uncharged matter it was shown that
phase-space anisotropies at equilibrium arise due to specifically-kinetic
effects associated with nonuniform gravitational fields together with
phase-space conservation laws. These include the conservation of angular
momentum in spherically-symmetric \cite{gravi,r85} or axially-symmetric
configurations \cite{IJMPD}, or the existence of Killing tensor symmetries
for the particle geodesic motion. In this regard, the statistical
description of equilibrium collisionless $N-$body systems in the Kerr
space-time has been proposed for the first time in Ref.\cite{ijmpd-2016},
where the existence of an intrinsically-relativistic mechanism yielding
non-isotropic KDFs and induced by the existence of the Carter constant for
single-particle dynamics has been proved.

Based on these considerations the target of the present study deals with the
investigation of the physical properties of non-isotropic kinetic equilibria
and fluid stress-energy tensors generated by Killing tensor symmetry in
curved space-time for uncharged matter. The astrophysical system considered
here is formed by a central rotating compact object and a surrounding
distribution of relativistic collisionless neutral particles, as could be
the realistic case of dark matter halos around rotating black holes. The
purpose is the address in further detail of the Carter-constant induced
kinetic mechanism first proposed in Ref.\cite{ijmpd-2016} and the new
solution obtained there for the equilibrium KDF differing from a Maxwellian
distribution, investigating possible relevant physical consequences on
astrophysical systems and their macroscopic observable features. In
particular, in this work we aim to abandon the assumption of having a test
fluid set over a prescribed background metric tensor, by consistently
including the contribution of its self-generated gravitational field. The
result is obtained in the framework of covariant kinetic theory by solving
the Einstein field equations in terms of an analytical perturbative
treatment. As such, the gravitational field is decomposed as a prescribed
background metric tensor generated by a rotating black hole and described by
the Kerr solution plus a self-field correction originating from the fluid
neutral matter in equilibrium configuration. The latter one satisfies the
linearized Einstein equations having the non-isotropic stress-energy tensor
computed in Ref.\cite{ijmpd-2016} as source term. It is then pointed out
that, given validity of this kinetic solution, the collisionless $N-$body
systems of neutral matter is able to generate at equilibrium a Kerr-like
metric tensor, so that the self-generated gravitational field reproduces the
qualitative features of the background one, e.g. in particular exhibiting
off-diagonal terms. For this reason, the kinetic mechanism reported here is
referred to as magnification effect of Kerr metric. The existence of this
solution is expected to be significant in view of the peculiar features of
the Kerr metric, like the existence of ergoregion and frame-dragging
effects, both on single-particle dynamics \cite{dan0,dan1,dan1b,dan2} as
well as matter accreting phenomena \cite{dan3a,dan3,dan4,dan5,dan5b,dan6}.

The structure of the paper is as follows. In Section 2 the phenomenon of
temperature anisotropy occurring in magnetized plasmas and induced by
magnetic moment conservation is recalled, which is useful for comparison
with the case of neutral matter. In Section 3 the anisotropy character of
the kinetic solution induced by Killing-tensor symmetry for neutral matter
in the field of rotating Kerr black hole is illustrated, while the
corresponding non-isotropic stress-energy tensor and its qualitative
features are discussed in Section 4. Section 5 investigates the properties
of the self gravitational field, proving that the self metric is of
Kerr-like type and proving the existence of the phenomenon of Kerr-metric
magnification. Concluding remarks are reported in Section 6.

\section{2 - \textbf{Temperature anisotropy in plasmas}}

From a statistical point of view, the concept of temperature anisotropy
refers to the occurrence of an anisotropy in the directional particle
velocity dispersions averaged over the KDF and measured with respect to a
given reference system. In collisionless astrophysical plasmas different
sources of temperature anisotropy have been pointed out at equilibrium,
which combine the contributions arising from single-particle dynamics and
non-uniform background gravitational and EM fields \cite%
{Cr2010,Cr2011,Cr2013,Cr2013b,Cr2013c,PRE-new}. The most relevant one is a
magnetic-related effect induced by the conservation of particle magnetic
moment as predicted by gyrokinetic (GK) theory \cite{Little83}. The reason
is that this mechanism relies on the velocity-space symmetry of the Larmor
rotation motion, and therefore it is independent of the existence of
additional space-time symmetries \cite{PoP2014-1,PoP2014-2}. In the
following we recall fundamental results of GK theory which are needed for
the present research, while a detailed treatment of its variational
formulation can be found in Refs.\cite{Bek1,Bek2,PoP2014-1,PoP2014-2} for
relativistic particle dynamics and in Ref.\cite{Cr2013} for the
corresponding non-relativistic limit.

The non-perturbative formulation of covariant GK theory is achieved in terms
of a guiding-center transformation of particle state of the form%
\begin{eqnarray}
r^{\mu } &=&r^{\prime \mu }+\rho _{1}^{\prime \mu },  \label{1} \\
u^{\mu } &=&u^{\prime \mu }\oplus \nu _{1}^{\prime \mu },  \label{2}
\end{eqnarray}%
where $r^{\prime \mu }$ is the guiding-center position $4-$vector, $\rho
_{1}^{\prime \mu }$ is referred to as the relativistic Larmor $4-$vector, $%
\oplus $ denotes the relativistic composition law which warrants that $%
u^{\mu }$ is a $4-$velocity and primed quantities are all evaluated at $%
r^{\prime \mu }$. A background EM field described by the antisymmetric
Faraday tensor $F_{\mu \nu }$ is assumed. Denoting with $H$\ and $E$\ the $%
4- $scalar eigenvalues of $F_{\mu \nu }$, the latter can be conveniently
represented as%
\begin{equation}
F_{\mu \nu }=H\left( c_{\nu }d_{\mu }-c_{\mu }d_{\nu }\right) +E\left(
b_{\mu }a_{\nu }-b_{\nu }a_{\mu }\right) .  \label{tetrad-fmunu}
\end{equation}%
which determines the orientation of the tetrad basis (EM-tetrad frame). The
physical meaning is that $H$\ and $E$\ coincide with the observable magnetic
and electric field strengths in the reference frame where the electric and
the magnetic fields are parallel. Notice that the EM-tetrad basis represents
the natural covariant generalization of the magnetic-related triad system
formed by the orthogonal right-handed unit 3-vectors $\left( \mathbf{e}_{1},%
\mathbf{e}_{2},\mathbf{e}_{3}\equiv \mathbf{b}\right) $\textbf{\ }usually
introduced in non-relativistic treatments \cite{Cr2013}, where $\mathbf{b}$
is the unit vector parallel to the local direction of the magnetic field and
$\left( \mathbf{e}_{1},\mathbf{e}_{2}\right) $ are two unit orthogonal
vectors in the normal plane to $\mathbf{b}$. When the $4-$vector $u^{\prime
\mu }$ is projected on the guiding-center EM-basis, it determines the
representation%
\begin{equation}
u^{\prime \mu }\equiv u_{0}^{\prime }a^{\prime \mu }+u_{\parallel }^{\prime
}b^{\prime \mu }+w^{\prime }\left[ c^{\prime \mu }\cos \phi ^{\prime
}+d^{\prime \mu }\sin \phi ^{\prime }\right] ,  \label{u-primo}
\end{equation}%
which defines the gyrophase angle $\phi ^{\prime }$ associated with the
Larmor rotation, where $w^{\prime }$ is the magnitude of $u^{\prime \mu }$
in the plane $\pi _{\perp }\equiv (c^{\prime \mu },d^{\prime \mu })$. Then,
denoting $\left\langle {}\right\rangle _{\phi ^{\prime }}\equiv \frac{1}{%
2\pi }\oint d\phi ^{\prime }$ the gyrophase-averaging operator, the
following non-perturbative representation is obtained for the relativistic
particle magnetic moment $m^{\prime }$:%
\begin{equation}
m^{\prime }=\left\langle \frac{\partial \rho _{1}^{\prime \mu }}{\partial
\phi ^{\prime }}\left[ \left( u_{\mu }^{\prime }\oplus \nu _{1\mu }^{\prime
}\right) +qA_{\mu }\right] \right\rangle _{\phi ^{\prime }},  \label{m-exact}
\end{equation}%
where $q\equiv \frac{Ze}{mc^{2}}$, $Ze$ is the particle electric charge and $%
A_{\mu }\left( r\right) $ is the EM $4-$potential.

Based on the exact GK solution, a covariant perturbative theory can be
developed for the analytical asymptotic treatment of Eq.(\ref{m-exact}),
yielding a perturbative representation of $m^{\prime }$. This is achieved by
a Larmor-radius expansion of the relevant dynamical quantities in terms of
the $4-$scalar dimensionless parameter $\varepsilon \equiv \frac{r_{L}}{L}$
to be assumed infinitesimal, namely $\varepsilon \ll 1$. Here $r_{L}\equiv
\sqrt{\rho _{1}^{\prime \mu }\rho _{1\mu }^{\prime }}$, while $L$ denotes a
characteristic invariant scale-length associated with the background fields
and to be properly defined. A plasma characterized by the asymptotic
ordering $\varepsilon \ll 1$\ is referred to as strongly-magnetized. It
means that its physical properties change on scale-lengths greater than the
Larmor radius. Upon carrying out the perturbative expansion to first order
in $\varepsilon $, one obtains that in the perturbative theory the $4-$%
vector $u^{\prime \mu }$ in Eq.(\ref{u-primo}) represents the leading-order
particle $4-$velocity. Similarly, the asymptotic representation $m^{\prime
}=\mu ^{\prime }\left[ 1+O\left( \varepsilon \right) \right] $ is reached,
where $\mu ^{\prime }=\frac{w^{\prime 2}}{2qH^{\prime }}$ is the
leading-order contribution of the magnetic moment which represents an
adiabatic invariant, with $H^{\prime }$ denoting the corresponding
guiding-center eigenvalue of $H$.

This result carries important physical consequences. In fact, in the
asymptotic formulation of GK theory, thanks to the conservation of $\mu
^{\prime }$, particle dynamics in the plane $\pi _{\perp }\equiv (c^{\prime
\mu },d^{\prime \mu })$ orthogonal to the local direction of magnetic field
decouples from that in the plane $\pi _{\parallel }\equiv (a^{\prime \mu
},b^{\prime \mu })$ containing the parallel space direction to it. This
feature is recovered also in the non-relativistic regime of GK dynamics
between the direction $\mathbf{b}$ and the orthogonal plane $\left( \mathbf{e%
}_{1},\mathbf{e}_{2}\right) $. The Larmor circular motion of single charges
in the plane $\pi _{\perp }$ is responsible for the emission of
elliptically-polarized synchrotron radiation. When averaged over a
distribution of particles in a magnetized plasma the resulting emission
retains a characteristic degree of linear polarization \cite{ryb}.

The functional form of the equilibrium KDF describing collisionless
magnetized plasmas must generally contain the particle magnetic moment.
Since its leading-order expression depends only on $w^{\prime 2}$ and not on
the remaining component $u_{\parallel }^{\prime }$ it follows that,
according to the definitions introduced above, $\mu ^{\prime }$ becomes
necessarily a source of phase-space anisotropy for the kinetic equilibrium,
producing a non-isotropic KDF \cite{Cr2010,Cr2011,Cr2013,PoP2014-1,PoP2014-2}%
. This generates a corresponding non-isotropic stress-energy tensor and the
statistical phenomenon of temperature anisotropy defined with respect to the
planes $\pi _{\perp }$ and $\pi _{\parallel }$. In analogy with the
synchrotron radiation emission, we denote this characteristic feature of
collisionless plasmas as \emph{polarization of temperature anisotropy},
implying that the thermodynamical properties in the plane $\pi _{\perp }$
and along the parallel direction $\mathbf{b}$ decouple. Finally, we notice
that the orientations of $\pi _{\perp }$ and $\mathbf{b}$ determined by the
EM-tetrad frame with respect to a distant fixed observer frame change in
configuration space according to the profile of the non-uniform EM field of
the system.

\section{3 - Anisotropy induced by \textbf{Killing-tensor symmetry}}

In this section we discuss the mechanism of generation of equilibrium
phase-space anisotropies induced by Killing-tensor symmetries. The case of a
collisionless $N-$body system composed of neutral matter is considered,
which is treated as a system testing the influence of a black-hole
space-time. In this approximation the Einstein and Vlasov equations
decouple, allowing for an analytical construction of kinetic equilibria in
a\ given background metric tensor.

We consider space-time solutions for $g_{\mu \nu }\left( r^{\alpha }\right) $
characterized by symmetry properties, namely admitting Killing tensor fields
\cite{wald}. This means that each Killing vector $C^{\mu }$ associated with
a cyclic coordinate generates the $4-$scalar integral of motion $C\equiv
C^{\mu }u_{\mu }$ which is linear in the particle $4-$velocity, while each
Killing tensor $K_{\mu \nu }$ associated with so-called hidden symmetries
generates the invariant integral of motion $K\equiv K_{\mu \nu }u^{\mu
}u^{\nu }$ which is quadratic in $u^{\mu }$. For an illustration of the
theory, the case of an uncharged rotating black hole of rest mass $M$ and
intrinsic angular momentum $S$ described by the Kerr metric is discussed.
Adopting Boyer-Lindquist coordinates $\left( t,\phi ,r,\theta \right) $ and
the geometrical system of units ($c=G=1$), the corresponding metric is
expressed in terms of the line element $ds$ as%
\begin{equation}
ds^{2}=-\left[ \frac{\Delta -a^{2}\sin ^{2}\theta }{\Sigma }\right] dt^{2}-%
\frac{4Mar\sin ^{2}\theta }{\Sigma }dtd\phi +\left[ \frac{\left(
r^{2}+a^{2}\right) ^{2}-\Delta a^{2}\sin ^{2}\theta }{\Sigma }\right] \sin
^{2}\theta d\phi ^{2}+\frac{\Sigma }{\Delta }dr^{2}+\Sigma d\theta ^{2},
\label{kerr}
\end{equation}%
where as usual $\Sigma \equiv r^{2}+a^{2}\cos ^{2}\theta $ and $\Delta
\equiv r^{2}+a^{2}-2Mr$, with $a\equiv S/M$ being the angular momentum per
unit mass. Correspondingly, in the Boyer-Lindquist coordinate system the
particle $4-$velocity is represented in terms of the unit $4-$vectors $%
\left( a^{t},b^{\phi },c^{r},d^{\theta }\right) $ defined along the
space-time directions identified by the same coordinate system as%
\begin{equation}
u^{\mu }\equiv u_{t}a^{t}+u_{\phi }b^{\phi }+u_{r}c^{r}+u_{\theta }d^{\theta
},  \label{u-bl}
\end{equation}%
so that the velocity components entering Eq.(\ref{4-veeldeco}) are now
denoted as $\left( u_{0},u_{1},u_{2},u_{3}\right) \equiv \left(
u_{t},u_{\phi },u_{r},u_{\theta }\right) $. Concerning the notation, it must
be noticed that in the previous equation the tensorial indices of the
controvariant $4-$vector $u^{\mu }$ are identified by the set of unit $4-$%
vectors $\left( a^{t},b^{\phi },c^{r},d^{\theta }\right) $, while the $4-$%
scalars $\left( u_{t},u_{\phi },u_{r},u_{\theta }\right) $\ denote only the
magnitude of the single directional components and should not be interpreted
as covariant tensors (i.e., the subscripts here are only labels that
identify the correct directional components of the $4-$velocity).

In the Kerr solution the coordinates $t$ and $\phi $ are ignorable, implying
existence of the two Killing vectors $\xi ^{\alpha }=\left( \partial
/\partial t\right) ^{\alpha }$ and $\zeta ^{\alpha }=\left( \partial
/\partial \phi \right) ^{\alpha }$. The corresponding integrals of motion
identify respectively the total particle energy $E$ and angular momentum $L$%
, which are written per unit rest mass for particle geodesic as%
\begin{equation}
E\equiv -\xi ^{\mu }u_{\mu }=\left[ 1-\frac{2Mr}{\Sigma }\right] \overset{%
\cdot }{t}+\frac{2Mar\sin ^{2}\theta }{\Sigma }\overset{\cdot }{\phi },
\label{E}
\end{equation}%
\begin{equation}
L\equiv \zeta ^{\mu }u_{\mu }=-\frac{2Mar\sin ^{2}\theta }{\Sigma }\overset{%
\cdot }{t}+\frac{\left( r^{2}+a^{2}\right) ^{2}-\Delta a^{2}\sin ^{2}\theta
}{\Sigma }\sin ^{2}\theta \overset{\cdot }{\phi },  \label{L}
\end{equation}%
where $\overset{\cdot }{t}=\frac{dt}{ds}$ and $\overset{\cdot }{\phi }=\frac{%
d\phi }{ds}$. In addition, the Kerr metric admits a Killing tensor related
to the angular momentum of the field source and generating the following
integral of the geodesic motion:%
\begin{equation}
K\equiv \mathcal{Q}+\left( L-aE\right) ^{2},  \label{K}
\end{equation}%
where $\mathcal{Q}$ denotes the \textit{Carter constant }\cite{Carter}%
\begin{equation}
\mathcal{Q=}p_{\theta }^{2}+\cos ^{2}\theta \left[ a^{2}\left(
1-E^{2}\right) +\left( \frac{L}{\sin \theta }\right) ^{2}\right] ,  \label{Q}
\end{equation}%
and here $p_{\theta }=\Sigma \overset{\cdot }{\theta }$, with $\overset{%
\cdot }{\theta }=\frac{d\theta }{ds}$. The scalar $K$ has the property of
being always non-negative \cite{gravi}. For the following developments, it
is useful to represent the invariant $K$ as a polynomial of the particle
velocity components as%
\begin{equation}
K=\Sigma ^{2}u_{\theta }^{2}+A_{1}u_{t}^{2}+A_{2}u_{\phi
}^{2}+A_{3}u_{t}u_{\phi }+A_{4},  \label{k-poli}
\end{equation}%
where $A_{1}$, $A_{2}$, $A_{3}$ and $A_{4}$ are configuration-space
functions, whose expression follows from Eqs.(\ref{E})-(\ref{K}) and are
given respectively by%
\begin{eqnarray}
A_{1} &\equiv &a^{2}\sin ^{2}\theta , \\
A_{2} &\equiv &\left( r^{2}+a^{2}\right) ^{2}\sin ^{2}\theta , \\
A_{3} &\equiv &-2a\left( r^{2}+a^{2}\right) \sin ^{2}\theta , \\
A_{4} &\equiv &a^{2}\cos ^{2}\theta .
\end{eqnarray}%
Finally, the last condition for having integrable motion is provided by Eq.(%
\ref{u-zero}), which becomes in Boyer-Lindquist notation%
\begin{equation}
u_{t}=\sqrt{u_{\phi }^{2}+u_{r}^{2}+u_{\theta }^{2}-1}.  \label{uut}
\end{equation}

Given the set of integrals of motion we can now consider the problem of
determining relativistic kinetic equilibria for the system of interest here.
The method of invariant is implemented, by identifying the integrals of
motion with the set $I\left( \mathbf{x}\left( s\right) \right) =(E,L,K)$
defined respectively by Eqs.(\ref{E}), (\ref{L}) and (\ref{K}). The
equilibrium KDF is represented in the form $f=f_{\ast }$, where%
\begin{equation}
f_{\ast }=f_{\ast }\left( \left( E,L,K\right) ,\Lambda _{\ast }\right)
\label{f-star}
\end{equation}%
is a smooth strictly-positive function of the particle invariants only,
which is summable in velocity-space. In Eq.(\ref{f-star}) $\left(
E,L,K\right) $ denotes the explicit functional dependence of $f_{\ast }$,
while $\Lambda _{\ast }$ represents set of structure functions \cite{Cr2011}%
, namely invariant functions which are suitably related to the observable
velocity moments of the KDF and to be properly defined. By construction, the
structure functions carry implicit functional dependences on the same set of
invariants. Their general representation is necessarily of the type $\Lambda
_{\ast }=\Lambda _{\ast }\left( E,L,K\right) $. A particular realization
corresponds to the case in which $\Lambda _{\ast }$ are identically
constant, namely $\Lambda _{\ast }=const.$

Following Ref.\cite{ijmpd-2016} it is possible to introduce an explicit
representation for the equilibrium KDF consistent with Eq.(\ref{f-star}) in
terms of a Gaussian-like distribution of the form%
\begin{equation}
f_{\ast }=\beta _{\ast }e^{-E\gamma _{\ast }-L\omega _{\ast }-K\alpha _{\ast
}}.  \label{equil-gaussian}
\end{equation}%
Here the ensemble $\Lambda _{\ast }=\left( \beta _{\ast },\gamma _{\ast
},\omega _{\ast },\alpha _{\ast }\right) $ identifies the set of invariant
structure functions, which are related to the system physical observables.
In particular, $\left( \beta _{\ast },\gamma _{\ast },\omega _{\ast }\right)
$ are equilibrium fields which arise in all Gaussian-like distributions
describing collisionless systems in stationary and axisymmetric
configurations, whose KDF depends on the conserved single-particle energy
and carries an azimuthal flow \cite{Cr2010,Cr2012,Cr2011,Cr2011a}. More
precisely, $\beta _{\ast }$ is associated with the system number density
measured in the fluid comoving frame, $\gamma _{\ast }$ determines the
system isotropic temperature, while $\omega _{\ast }$ enters the definition
of the fluid angular frequency in the $\phi $-direction when measured by an
inertial observer. In contrast, $\alpha _{\ast }$ is characteristic of the
anisotropy character of the solution associated with $K$ and can be shown to
be related to the system temperature anisotropy. This is a consequence of
the non-isotropic dependence of the constant $K$ on the directional
components of single-particle velocity (see Eq.(\ref{Q})). The role of the
invariant $K$ in the KDF (\ref{equil-gaussian}) is analogous to that played
by the particle magnetic moment $m^{\prime }$ in relativistic plasmas, so
that the contribution $K\alpha _{\ast }$ in $f_{\ast }$ effectively
generates a deviation from the isotropic solution. This provides a source of
phase-space anisotropy induced by Killing tensor symmetry. In fact, the same
invariant $K$ is a non-isotropic function of the squared of the velocity
components $u_{\theta }$, $u_{t}$ and $u_{\phi }$, so that its inclusion in
Eq.(\ref{equil-gaussian}) naturally generates a non-isotropic equilibrium
KDF.

In order to investigate the functional role of the Carter-constant $\mathcal{%
Q}$ in the effect of magnification of Kerr metric to be studied below,
without loss of generality we shall assume heron that all the structure
functions appearing in Eq.(\ref{equil-gaussian}) are identically constant,
namely that $\Lambda _{\ast }=const.$ For consistency with the notation
introduced in Ref.\cite{ijmpd-2016} we shall denote the set of constant
structure functions as $\Lambda =\left( \beta ,\gamma ,\omega ,\alpha
\right) $. This choice permits us to single out the explicit functional
dependence of $f_{\ast }$ on $K$, ignoring possible implicit dependences
that would require setting up an appropriate perturbative treatment
analogous to that developed in Refs.\cite{Cr2010,Cr2011,Cr2011a}. It must be
noticed that the condition $\Lambda _{\ast }=const.$ does not imply that the
observable fluid fields of the system associated with $f_{\ast }$ are
constant too, since the equilibrium KDF still retains configuration-space
dependences which are due to the non-uniform background metric tensor and
contribute through the same invariants $\left( E,L,K\right) $. These type of
spatial dependences are in fact inherited by the continuum fluid fields once
integration of the KDF over the velocity space is performed.

In view of these assumptions we then notice that Eq.(\ref{equil-gaussian})
can be equivalently written as%
\begin{equation}
f_{\ast }=f_{M}\Delta f,  \label{f-seconda}
\end{equation}%
where respectively%
\begin{equation}
f_{M}\equiv \beta e^{-E\gamma -L\omega }
\end{equation}%
identifies the isotropic Maxwellian distribution carrying
spatially-non-uniform number density, temperature and bulk flow velocity,
while%
\begin{equation}
\Delta f\equiv e^{-K\alpha }
\end{equation}%
denotes the non-isotropic correction factor due specifically to the
phase-space constraint placed on the equilibrium distribution $f_{\ast }$ by
the Killing-tensor conservation law. The quantity $\Delta f$ is in fact
responsible for the generation of non-isotropic stress-energy tensor (see
discussion below).

Before concluding this section, it is worth further elaborating the solution
(\ref{f-seconda}) to make it analytically tractable for the scope of this
paper. This is obtained introducing the so-called \emph{weak-anisotropy
regime} pointed out in Ref.\cite{ijmpd-2016}. More precisely, let us assume
that the ordering assumption%
\begin{equation}
K\alpha \sim O\left( \varepsilon \right)  \label{war}
\end{equation}%
holds in a suitable subset of phase-space for a proper prescription of the
constant $\alpha $, where $\varepsilon \ll 1$ is a dimensionless parameter
of the system that can be related for example to the scale of change of
fluid fields compared to spatial variation of the background metric tensor
(for a proper mathematical definition see Ref.\cite{ijmpd-2016}). The
condition (\ref{war}) corresponds to a configuration in which the
contribution of the invariant $K$ to the equilibrium solution carried by $%
\Delta f$ is of higher-order in $\varepsilon $ with respect to the other
invariants contained in $f_{M}$ and allows for a treatment of non-isotropic
stress-energy tensor in a perturbative analytical way. When Eq.(\ref{war})
is met, the equilibrium KDF $f_{\ast }$ in Eq.(\ref{f-seconda}) can be
Taylor-expanded to first-order in $\varepsilon $ to give%
\begin{equation}
f_{\ast }=f_{w}\left[ 1+O\left( \varepsilon ^{2}\right) \right] ,
\end{equation}%
where $f_{w}$ denotes the distribution%
\begin{equation}
f_{w}=f_{M}\left( 1-K\alpha \right) .  \label{war-kdf}
\end{equation}

From the physical point of view, in the weak-anisotropy regime considered
here the anisotropic character of the equilibrium kinetic solution is
uniquely associated with the term $K\alpha $ that acts as a polynomial of
second degree on particle velocity and provides a first-order correction to
the leading-order isotropic Maxwellian distribution $f_{M}$. Since the
invariant $K$ is actually a peculiar feature of a rotating black-hole
space-time, described here by the Kerr solution, it follows that the
conceptual difference between a pure isotropic Maxwellian solution $f_{M}$
and the non-isotropic one $f_{w}$\ is relevant and non-negligible, since $%
f_{w}$ includes all the peculiarities of the curved space-time solution for
rotating black-hole and related particle dynamics in General Relativity.

\section{4 - Non-isotropic stress-energy tensor}

In this section the qualitative properties of the stress-energy tensor
defined by Eq.(\ref{tmunu-bis}) and corresponding to the anisotropic
equilibrium KDF $f_{\ast }$ prescribed above are presented. Since the same
kinetic equilibrium for collisionless neutral matter is described by a
non-isotropic KDF, it follows that necessarily also the corresponding tensor
field $T^{\mu \nu }\left( r^{\alpha }\right) $ is non-isotropic, in the
sense defined in the Introduction. In principle, assuming the general case
of non-uniform structure functions $\Lambda _{\ast }$, the explicit
evaluation of the components of the tensor $T^{\mu \nu }\left( r^{\alpha
}\right) $ can be generally achieved by means of numerical integration, once
the preliminary prescriptions of the functional form of the same functions\ $%
\Lambda _{\ast }$ as well as the background metric tensor\ $g_{\mu \nu
}\left( r^{\alpha }\right) $ is done. Under these circumstances, it is
possible to conclude that the non-isotropic character of $T^{\mu \nu }\left(
r^{\alpha }\right) $ arises as a consequence of the combined action of the
dependence of $f_{\ast }$ on the invariant $K$ and the non-uniform
phase-space profiles of the structure functions, being both these features
characteristic of the relativistic equilibrium solution in the Kerr
background metric tensor.

However, for the purpose of the present work, we leave the numerical
analysis of $T^{\mu \nu }\left( r^{\alpha }\right) $ to future studies, and
we concentrate here on the analytical approach, which is sufficient to prove
the existence of the magnification effect of the Kerr metric mentioned
above. More precisely, this can be gained by considering the kinetic
equilibrium solution in the weak-anisotropy regime, whereby the
leading-order equilibrium KDF is represented by $f_{w}$ defined in Eq.(\ref%
{war-kdf}). Thus, in this\ regime, ignoring corrections of\ $O\left(
\varepsilon ^{2}\right) $, the stress-energy tensor becomes%
\begin{equation}
T^{\mu \nu }\left( r^{\alpha }\right) =\int \frac{\sqrt{-g}d^{3}u}{\sqrt{%
u_{\phi }^{2}+u_{r}^{2}+u_{\theta }^{2}-1}}u^{\mu }u^{\nu }f_{M}\left(
1-K\alpha \right) ,  \label{T-weak}
\end{equation}%
where we have adopted Boyer-Lindquist coordinates and in the integral the
component $u_{t}$ is given by Eq.(\ref{u-zero}). Because of the polynomial
representation of the equilibrium KDF over the Maxwellian expression $f_{M}$%
, the tensor $T^{\mu \nu }\left( r^{\alpha }\right) $ in Eq.(\ref{T-weak})
can be equivalently decomposed as%
\begin{equation}
T^{\mu \nu }\left( r^{\alpha }\right) =T_{M}^{\mu \nu }\left( r^{\alpha
}\right) +\Pi ^{\mu \nu }\left( r^{\alpha }\right) .
\label{stress-energy-tensor-total}
\end{equation}%
Here, $T_{M}^{\mu \nu }\left( r^{\alpha }\right) $ represents the
leading-order contribution associated with the isotropic Maxwellian KDF $%
f_{M}$, which \textit{per se} describes ideal fluids. Its customary
tensorial form is written as%
\begin{equation}
T_{M}^{\mu \nu }\left( r^{\alpha }\right) =neU^{\mu }U^{\nu }-p\Delta ^{\mu
\nu },  \label{T-max}
\end{equation}%
where $n$ is the particle number density and $e$ is the energy per particle,
so that $ne$ is the fluid energy density, $U^{\mu }$ is the fluid $4-$%
velocity, $p$ is the scalar isotropic pressure and $\Delta ^{\mu \nu }$ is
the so-called projector operator $\Delta ^{\mu \nu }=g^{\mu \nu }-U^{\mu
}U^{\nu }$ (see Ref.\cite{degroot} for a comprehensive treatment of
relativistic Maxwellian fluid fields). At this order\ in this asymptotic
evaluation, no contribution arises from the invariant $K$, so that, in the
weak-anisotropy regime, the leading-order stress-energy tensor is isotropic.

In contrast, the non-isotropic character of $T^{\mu \nu }\left( r^{\alpha
}\right) $ is carried by the $O\left( \varepsilon \right) $ term $\Pi ^{\mu
\nu }\left( r^{\alpha }\right) $, which, from Eq.(\ref{T-weak}) and under
the condition $\alpha =const.$, is found to be a weighted\ integral on a
Maxwellian KDF of the form%
\begin{equation}
\Pi ^{\mu \nu }\left( r^{\alpha }\right) =-\alpha \int \frac{\sqrt{-g}d^{3}u%
}{\sqrt{u_{\phi }^{2}+u_{r}^{2}+u_{\theta }^{2}-1}}u^{\mu }u^{\nu }Kf_{M}.
\label{stress-ener}
\end{equation}%
In the weak-anisotropy regime, the non-isotropic character of $\Pi ^{\mu \nu
}\left( r^{\alpha }\right) $ arises only due to the dependence on the
invariant $K$. In fact, invoking the representation (\ref{k-poli}) given
above, one has that the integrand in the previous equation becomes a
polynomial function carrying a non-isotropic dependence on the particle
velocity components $\left( u_{\phi },u_{r},u_{\theta }\right) $.

The contributions arising from $\Pi ^{\mu \nu }\left( r^{\alpha }\right) $
can be evaluated explicitly by adopting Boyer-Lindquist coordinates and
representing the particle $4-$velocity from Eq.(\ref{4-veeldeco}) in terms
of the unit $4-$vectors $\left( a^{t},b^{\phi },c^{r},d^{\theta }\right) $ as%
\begin{equation}
u^{\mu }\equiv u_{t}a^{t}+u_{\phi }b^{\phi }+u_{r}c^{r}+u_{\theta }d^{\theta
}.
\end{equation}%
We remark that, once $u^{\mu }$ is represented in such a tetrad in terms of
the basis formed by $\left( a^{t},b^{\phi },c^{r},d^{\theta }\right) $, the
same $4-$vectors also identify the tensorial components of $T^{\mu \nu
}\left( r^{\alpha }\right) $, which are generally position-dependent.

Let us then consider the representation (\ref{k-poli}) for the invariant $K$%
. Manifestly, the integral on the configuration-space function $A_{4}$
yields simply a first-order isotropic contribution to the tensor $T_{M}^{\mu
\nu }\left( r^{\alpha }\right) $. The non-isotropic features arise therefore
from the terms proportional to $\Sigma ^{2}$, $A_{1}$, $A_{2}$ and $A_{3}$
in Eq.(\ref{k-poli}). We analyze first the diagonal components of $\Pi ^{\mu
\nu }\left( r^{\alpha }\right) $. In this reference, we notice that the term
proportional to $A_{3}$ does not contribute being odd in $u_{\phi }$. It
follows that the contribution $\Pi ^{tt}\left( r^{\alpha }\right) $ is given
by%
\begin{eqnarray}
\Pi ^{tt}\left( r^{\alpha }\right) &=&-\alpha a^{t}a^{t}\int \sqrt{-g}d^{3}u%
\sqrt{u_{\phi }^{2}+u_{r}^{2}+u_{\theta }^{2}-1}  \notag \\
&&\left[ \Sigma ^{2}u_{\theta }^{2}+A_{1}u_{t}^{2}+A_{2}u_{\phi }^{2}\right]
f_{M},  \label{pi-tt}
\end{eqnarray}%
while the component $\Pi ^{\phi \phi }\left( r^{\alpha }\right) $ is instead%
\begin{eqnarray}
\Pi ^{\phi \phi }\left( r^{\alpha }\right) &=&-\alpha b^{\phi }b^{\phi }\int
\frac{\sqrt{-g}d^{3}u}{\sqrt{u_{\phi }^{2}+u_{r}^{2}+u_{\theta }^{2}-1}}%
u_{\phi }^{2}  \notag \\
&&\left[ \Sigma ^{2}u_{\theta }^{2}+A_{1}u_{t}^{2}+A_{2}u_{\phi }^{2}\right]
f_{M}.  \label{pi-fifi}
\end{eqnarray}%
The remaining diagonal components $\Pi ^{rr}\left( r^{\alpha }\right) $ and $%
\Pi ^{\theta \theta }\left( r^{\alpha }\right) $ are formally analogous to $%
\Pi ^{\phi \phi }\left( r^{\alpha }\right) $ and can be obtained by
replacing in Eq.(\ref{pi-fifi}) the term $b^{\phi }b^{\phi }u_{\phi }^{2}$
respectively with $c^{r}c^{r}u_{r}^{2}$ and $d^{\theta }d^{\theta }u_{\theta
}^{2}$. It is immediate to conclude that all the diagonal terms of the
tensor are generally different from each other, the result of the
integration depending on the degree of the polynomial of the velocity
components over which the KDF $f_{M}$ is integrated.

Finally, contrary to the leading-order tensor $T_{M}^{\mu \nu }\left(
r^{\alpha }\right) $ which is purely diagonal, the first-order contribution $%
\Pi ^{\mu \nu }\left( r^{\alpha }\right) $ carries also non-vanishing
off-diagonal terms. These are the symmetric entries $\Pi ^{t\phi }\left(
r^{\alpha }\right) =\Pi ^{\phi t}\left( r^{\alpha }\right) $ which arise due
to the term $A_{3}u_{t}u_{\phi }$ in Eq.(\ref{k-poli}) and are given by%
\begin{equation}
\Pi ^{t\phi }\left( r^{\alpha }\right) =-\alpha a^{t}b^{\phi }A_{3}\int
\sqrt{-g}d^{3}u\sqrt{u_{\phi }^{2}+u_{r}^{2}+u_{\theta }^{2}-1}u_{\phi
}^{2}f_{M}.  \label{pi-ti-fi}
\end{equation}

A comparison of the expression obtained for $\Pi ^{\mu \nu }\left( r^{\alpha
}\right) $ with the representation given in Eq.(\ref{T-max}) for the
isotropic leading-order tensor $T_{M}^{\mu \nu }\left( r^{\alpha }\right) $
reveals the physical meaning of the non-isotropic character of the solution.
Indeed, the existence of the contribution $\Pi ^{t\phi }\left( r^{\alpha
}\right) $ shows that the consistent treatment of relativistic collisionless
equilibria in the curved space-time of a rotating black hole and the
inclusion of the invariant $K$ in the equilibrium KDF consistent with
Killing-tensor symmetry determine a non-trivial structure for the
stress-energy tensor, in comparison with the representation holding for
isotropic Maxwellian distributions of ideal fluids. In fact, a first kind of
anisotropy arises because the diagonal terms of the stress-energy tensor $%
\Pi ^{\mu \nu }\left( r^{\alpha }\right) $ (and therefore $T^{\mu \nu
}\left( r^{\alpha }\right) $) differ from each other. Second, the same
tensor is generally non-diagonal when expressed in the Boyer-Lindquist
coordinate system, with the non-vanishing symmetric terms provided by the
components $\Pi ^{t\phi }\left( r^{\alpha }\right) =\Pi ^{\phi t}\left(
r^{\alpha }\right) $ defined by the integral in Eq.(\ref{pi-ti-fi}).

In summary, from this analysis it follows that the matrix representation of
the stress-energy tensor given by Eq.(\ref{stress-energy-tensor-total}) in
the Boyer-Lindquist coordinates $\left( t,\phi ,r,\theta \right) $ is given
by%
\begin{equation}
T^{\mu \nu }\left( r^{\alpha }\right) =\left(
\begin{array}{cccc}
T_{M}^{tt}+\Pi ^{tt} & \Pi ^{\phi t} & 0 & 0 \\
\Pi ^{t\phi } & T_{M}^{\phi \phi }+\Pi ^{\phi \phi } & 0 & 0 \\
0 & 0 & T_{M}^{rr}+\Pi ^{rr} & 0 \\
0 & 0 & 0 & T_{M}^{\theta \theta }+\Pi ^{\theta \theta }%
\end{array}%
\right) .  \label{T-matrix}
\end{equation}

To conclude this section, it is useful to make a brief comparison of the
anisotropy effect pointed out here for relativistic collisionless neutral
matter with the one mentioned above holding for relativistic plasmas and
induced by magnetic moment conservation. Both types of anisotropy reflect
themselves in the non-isotropic character of the stress-energy tensor
obtained as a velocity moment of the equilibrium distribution function. This
tensor describes the observable thermal properties of the corresponding
continuum fluid system, namely its temperature and pressure profiles. Hence,
both astrophysical plasmas and neutral matter (e.g., collisionless dark
matter halos) characterized by non-isotropic stress-energy tensors are
expected to exhibit thermal properties which are non-isotropic in space,
with pressure and temperature profiles that vary according to the direction
of measurement. In the case of plasmas the system of reference with respect
to which the anisotropy is defined is determined by the local direction of
the magnetic field (see Eq.(\ref{u-primo})), while for neutral matter around
Kerr black-hole it is set by Boyer-Lindquist coordinates. Thus,
specifically, the origin of the anisotropy generated by $m^{\prime }-$%
conservation in plasmas (see Eq.(\ref{m-exact})) is purely magnetic, while
the Carter-constant induced one is purely geometric and is associated with
the properties of background space-time. Nevertheless, in both cases the
anisotropy is generated by some type of "subtle" but relevant symmetry,
which for plasmas is associated with a symmetry in velocity-space (gyrophase
motion of charges around magnetic field lines), while for Kerr black-hole is
a conservation law intrinsic of the background space-time, which is
quadratic in the single-particle velocity and is generated by Killing tensor
symmetry.

\section{5 - The self gravitational field: Kerr-like metric}

In this section we set up the mathematical framework for the treatment of
the gravitational field in an astrophysical system composed by a central
rotating black hole surrounded by relativistic collisionless neutral matter.
In fact, in Ref.\cite{ijmpd-2016} only the background gravitational field
was assumed, ignoring in first place the self-field generated by the
collisionless matter. The target of the present work is to proceed a step
further in the theory by determining the metric tensor contribution
generated by the neutral matter whose constituent particle dynamics is
determined by the background gravitational field generated by the central
compact object. Therefore, in the following we go beyond the assumption of
having a test fluid set over a prescribed field metric, and we determine the
corresponding self-generated gravitational field contribution. To reach the
goal the Einstein equations are solved by means of an analytical
perturbative treatment, whereby the gravitational field is decomposed as a
prescribed background metric tensor generated by the rotating Kerr black
hole and a self-field correction originating from the fluid neutral matter
at equilibrium and whose stress-energy tensor has been computed in Section 4.

To start with, we consider the Einstein field equations in absence of
cosmological constant%
\begin{equation}
\widehat{R}_{\mu \nu }-\frac{1}{2}\widehat{R}\widehat{g}_{\mu \nu }=8\pi
\widehat{T}_{\mu \nu },  \label{EINSTEIN FIELD EQS}
\end{equation}%
where as usual\textbf{\ }$\widehat{R}_{\mu \nu }$, $\widehat{R}\equiv
\widehat{g}^{\alpha \beta }\widehat{R}_{\alpha \beta }$ and $\widehat{T}%
_{\mu \nu }$ denote respectively the Ricci tensor, the Ricci scalar and the
total stress-energy tensor due to external sources, all evaluated in terms
of the metric tensor $\widehat{g}_{\mu \nu }$. Here $\widehat{g}_{\mu \nu }$
denotes the total metric tensor solution of the non-linear tensorial
equations (\ref{EINSTEIN FIELD EQS}) for appropriate initial/boundary
conditions and prescription of source terms through $\widehat{T}_{\mu \nu }$%
. In order to determine the contribution of the neutral matter to the system
metric tensor, we treat the corresponding stress-energy tensor in a
perturbative way, so that we introduce the tensor decomposition%
\begin{equation}
\widehat{g}_{\mu \nu }=g_{\mu \nu }+h_{\mu \nu },  \label{tensor-decomp}
\end{equation}%
where $g_{\mu \nu }$ is the leading-order contribution, corresponding to the
background metric tensor, while $h_{\mu \nu }$ is the first-order
perturbation.

Let us analyze the two terms on the rhs of Eq.(\ref{tensor-decomp})
separately, starting with $g_{\mu \nu }$. This represents the background
metric tensor which raises and lowers tensor indices and prescribes the
standard connections (Christoffel symbols) appearing in the covariant
derivatives. It is assumed that $g_{\mu \nu }$ satisfies the Einstein field
equations in which the only source term is provided by the mass distribution
due to the rotating compact object. As such, in the external region to the
event horizon $g_{\mu \nu }$ represents a vacuum field, namely the solution
of the Einstein field equations in vacuum subject to the boundary conditions
set on the black hole domain. In accordance with Ref.\cite{ijmpd-2016}, this
is taken to be a rotating Kerr black hole, so that the background metric
tensor can be expressed in terms of the line element $ds$ as in Eq.(\ref%
{kerr}) in terms of Boyer-Lindquist coordinates.

In difference, $h_{\mu \nu }$ represents the perturbation of the metric
tensor, which is assumed to be "small" compared to $g_{\mu \nu }$.
Therefore, as is customary done in the literature, under this assumption by
implementing the decomposition (\ref{tensor-decomp}) it is possible to
linearize the Einstein field equations to first-order in $h_{\mu \nu }$ and
determine the differential equation solving for the same tensor $h_{\mu \nu
} $. Thus, introducing the trace-reversed tensor $\gamma _{\mu \nu }$
defined as%
\begin{equation}
\gamma _{\mu \nu }=h_{\mu \nu }-\frac{1}{2}\left( g^{\alpha \beta }h_{\alpha
\beta }\right) g_{\mu \nu },  \label{trace-revrsed}
\end{equation}%
and imposing the Lorentz gauge condition $\nabla _{\nu }h^{\mu \nu }=0$, we
obtain the differential wave-like equation%
\begin{equation}
\square \gamma ^{\mu \nu }+2R_{\alpha \beta }^{\mu \nu }\gamma ^{\alpha
\beta }=-16\pi T^{\mu \nu }.  \label{wave eq}
\end{equation}%
Here, $\square \equiv \nabla _{\nu }\nabla ^{\nu }$ is the D'Alembertian
differential operator and $R_{\alpha \mu \beta \nu }$ the Riemann tensor,
both evaluated in terms of the background metric tensor $g_{\mu \nu }$. In
addition, on the rhs of the previous equation $T^{\mu \nu }$ denotes the
perturbing stress-energy tensor, which for the problem under investigation
here is associated with the distribution of collisionless neutral matter in
the external region of the rotating black hole. The formal solution of Eq.(%
\ref{wave eq}) for $\gamma ^{\mu \nu }=\gamma ^{\mu \nu }\left( r^{\alpha
}\right) $ can be written as%
\begin{equation}
\gamma ^{\mu \nu }\left( r^{\alpha }\right) =4\int \sqrt{-g^{\prime }}%
d^{4}r^{\prime }G_{\alpha ^{\prime }\beta ^{\prime }}^{\mu \nu }\left(
r^{\alpha },r^{\alpha \prime }\right) T^{\alpha ^{\prime }\beta ^{\prime
}}\left( r^{\alpha \prime }\right) ,  \label{solution-wave-eq}
\end{equation}%
where $G_{\alpha ^{\prime }\beta ^{\prime }}^{\mu \nu }\left( r^{\alpha
},r^{\alpha \prime }\right) $ is the Hadamard symmetric Green's function
(see Ref.\cite{poisson-lr}). Once $\gamma ^{\mu \nu }\left( r^{\alpha
}\right) $ is known, the metric perturbation $h_{\mu \nu }$ is obtained from
Eq.(\ref{trace-revrsed}). The perturbative treatment of the gravitational
field permits therefore the consistent inclusion of the contribution of the
self-field generated by the collisionless matter set in equilibrium
configuration under the influence of the central background gravitational
field. The complete information about the matter distribution around the
black hole is carried by the corresponding stress-energy tensor $T^{\mu \nu
} $ which has been previously evaluated in terms of the analytical solution
for the equilibrium KDF. We therefore expect that the properties of the
background metric tensor encoded in $T^{\mu \nu }$ through the kinetic
constraints associated with the Killing tensor symmetry and the existence of
the Carter constant should be inherited by the perturbative metric $h_{\mu
\nu }$ thanks to Eqs.(\ref{trace-revrsed}) and (\ref{solution-wave-eq}).

In order to elucidate the issue, we analyze the qualitative properties of
the solution (\ref{solution-wave-eq}) satisfying Eq.(\ref{wave eq}), where
the fluid stress-energy tensor is provided by Eq.(\ref%
{stress-energy-tensor-total}). From the analysis of the previous section we
know that, when expressed in Boyer-Lindquist coordinates, the stress-energy
tensor $T^{\mu \nu }$ takes the matrix representation (\ref{T-matrix}). The
discussion must include two separate possibilities, corresponding
respectively to the case in which the tensor components of $T^{\mu \nu }$
are null and the case in which they are different from zero (i.e., the
diagonal terms and the $t\phi -$terms). In particular:

A)\ For the tensorial components for which $T^{\mu \nu }=0$ in Eq.(\ref%
{T-matrix}), the differential wave equation (\ref{wave eq}) reduces to a
vacuum equation without source term. This provides a trivial solution for
the metric perturbation $h_{\mu \nu }$. In fact, the background metric
tensor $g_{\mu \nu }$ is a vacuum solution too, so that in the framework of
a perturbative theory the vacuum perturbation $h_{\mu \nu }$ does not
provide any additional information as it can be absorbed in $g_{\mu \nu }$
itself.

B)\ For tensorial components for which $T^{\mu \nu }\neq 0$ in Eq.(\ref%
{T-matrix}), the differential wave equation (\ref{wave eq}) is indeed a
non-vacuum equation with prescribed source term. For the tensorial indices
for which this occurs, its solution yields a modification of the background
metric tensor, consistent with the decomposition (\ref{tensor-decomp}).
However, we notice that the non-vanishing entries of $T^{\mu \nu }$ coincide
with the non-vanishing components of the background metric tensor in
Boyer-Lindquist coordinates, including the off-diagonal terms so relevant
for frame-dragging effects. Therefore, the immediate consequence is that in
the present perturbative framework, the analytical evaluation of the
self-generated gravitational field yields corrections to the Kerr solution
preserving the structural form of the metric tensor. In other words, the
self-generated field metric tensor is of Kerr-like type, resembling the same
qualitative features of the background metric tensor. This proves that when
a collisionless relativistic system of neutral particles sets in equilibrium
configuration in the background gravitational field of a rotating Kerr black
hole, the perturbation that it generates to the total system metric tensor
keeps memory of the background one, affecting all its component
representations. We denote this phenomenon \emph{magnification of Kerr metric%
}.

In summary, from this discussion one has that the matrix representation of
the self-metric $h_{\mu \nu }$ in the Boyer-Lindquist coordinates $\left(
t,\phi ,r,\theta \right) $ is given by%
\begin{equation}
h^{\mu \nu }\left( r^{\alpha }\right) =\left(
\begin{array}{cccc}
h^{tt} & h^{\phi t} & 0 & 0 \\
h^{t\phi } & h^{\phi \phi } & 0 & 0 \\
0 & 0 & h^{rr} & 0 \\
0 & 0 & 0 & h^{\theta \theta }%
\end{array}%
\right) .  \label{h-matrix}
\end{equation}%
The precise calculation of the single entries require implementing numerical
integration of the double integrals leading respectively first to $T^{\mu
\nu }$ and then to $\gamma ^{\mu \nu }$. This goes beyond the scope of the
present paper and is left to future dedicated studies. Here the relevant
outcome concerns the proof of the validity of the representation (\ref%
{h-matrix}) and the understanding of its physical properties and
implications.

An important remark is in order concerning this conclusion. In fact, the
existence of the magnification effect pointed out here is a unique
consequence of the general kinetic solution discovered in Ref.\cite%
{ijmpd-2016}, which takes into account in the statistical description of
collisionless neutral matter all the kinematic constraints to
single-particle particle dynamics that arise in Kerr metric, and in
particular the Carter-constant integral of motion. We notice in fact that
\textit{a priori} it is not obvious that the perturbed metric due to the
matter distribution keeps the same structure of the background one. Indeed,
if one ignores the constraint placed by Killing tensor symmetry, it is
possible to construct kinetic equilibria which are at most of Maxwellian
type, and therefore corresponding to isotropic solutions. For Maxwellian
distributions the stress-energy tensor $T^{\mu \nu }$ coincides with $%
T_{M}^{\mu \nu }$, which can generate only diagonal self-metric tensors.
Hence, for systems in the field of a rotating black hole, isotropic
Maxwellian solutions are intrinsically non-complete, as they ignore the
contribution of the Killing tensor symmetry. As a consequence, similar to
what happens for collisionless plasmas, the choice of isotropic Maxwellian
distributions commonly invoked in the literature ignores \textit{a priori}
the complex phenomenology that characterize the non-isotropic equilibrium
configuration discovered in Ref.\cite{ijmpd-2016}.

\section{6 - Conclusions}

In this paper the study of the physical properties associated with
astrophysical systems composed by a rotating central compact object
identified with a Kerr black hole endowed with a distribution of
relativistic collisionless neutral particles has been reported. In
particular, focus has been given to the qualitative features of the
self-generated metric tensor associated with the same neutral matter, and
treated here as a perturbation of the background metric tensor generated by
the compact object itself. Concrete applications of the model considered
here may include dark matter halos or gas clouds surrounding rotating black
holes, while the results are relevant for characterizing the properties of
the corresponding gravitational field metric tensor and, in turn, also the
dynamics of additional components that can be found in these systems, like
for example astrophysical plasmas. In particular, here the discovery of a
phenomenon referred to as \emph{magnification of Kerr metric} has been
reported. This refers to the proof that the self-generated metric tensor by
the neutral matter particles shares the same qualitative features of the
background metric tensor due to the rotating central black hole. In
Boyer-Lindquist coordinates this means that the non-vanishing components of
the self-metric are the same of the background metric tensor, so that in
particular the self-metric exhibits non-null off-diagonal terms that
reproduce those of the background Kerr metric. The resulting total metric
tensor of the system is therefore expected to exhibit a Kerr-like metric
magnified by the presence of collisionless matter, including the notorious
drag-frame effects associated with off-diagonal metric components. The
outcome is not obvious \textit{a priori} to hold. The result has been
obtained in the framework of relativistic kinetic theory by implementing
analytical calculation. More precisely, to reach the conclusion the proper
identification of the equilibrium configurations of collisionless neutral
matter setting up in the gravitational field of rotating black hole has been
obtained. It has been shown that the general solution for the equilibrium
kinetic distribution function must take account of the constraints placed on
single-particle dynamics by the existence of Killing tensor symmetry, namely
the so-called Carter-constant integral of motion. When this is done, the
resulting statistical equilibrium exhibits the characteristic property of
being non-isotropic in velocity space, thus differing from the customary
isotropic Maxwellian distribution. The implication is that also the
corresponding stress-energy tensor becomes non-isotropic, permitting to
infer that these systems should be intrinsically characterized by
non-isotropic thermal properties, and more precisely the occurrence of
temperature and pressure anisotropies. The same physical mechanism has been
finally found to be the root of the magnification effect of the Kerr metric,
while isotropic Maxwellian solutions that ignore the constraints placed by
Killing tensor symmetry cannot treat consistently this effect and its
phenomenology.

The theoretical framework proposed for the statistical description of the
dynamical and thermodynamical properties of collisionless neutral matter in
relativistic astrophysical regimes and the investigation of the
corresponding non-uniform space-time metric generation are relevant as they
permit to prove the existence of unexpected physical phenomena, like the
Kerr metric magnification. It is believed that the kinetic theory developed
here and the perturbative analytical calculation proposed for the evaluation
of the matter self metric might also provide the basis for future numerical
studies of the problem in concrete astrophysical systems.

\bigskip

\textbf{Acknowledgments - }Work developed within the research project of the
Albert Einstein Center for Gravitation and Astrophysics, Czech Science
Foundation No. 14-37086G (Z.S.).

\end{document}